\documentclass[a4paper]{jpconf}
\usepackage{graphicx}
\usepackage{slashed}
\usepackage{amsmath}
\usepackage{amsfonts}
\usepackage{amssymb}
\newcommand{\RCT}{R{$\chi$}T}
\newcommand{\etap}{\eta^{(\prime)}}
\newcommand{\deceta}{\tau\to\pi\eta\gamma\nu_\tau}
\newcommand{\decetap}{\tau\to\pi\eta'\gamma\nu_\tau}
\newcommand{\MeV}{\text{ MeV}}

\begin{document}
\title{Backgrounds in the search for Second Class Currents in $\tau$ decays}

\author{Adolfo Guevara}

\address{Departamento de F\'isica, Cinvestav IPN, Apdo. Postal 14-740, 07000 Ciudad de M\'exico, M\'exico}

\ead{aguevara@fis.cinvestav.mx}

\begin{abstract}
As constructed, the Standard Model does not include genuine Second Class Currents, however these can be induced
through the breaking of isospin or charge conjugation. The experimental limits from B-Factories are getting closer to the predictions made for
induced Second Class Currents through isospin breaking in the processes $\tau\to\pi\etap\nu_\tau$, therefore a more careful analysis of the
background in these decays becomes necessary. In this work we analyze the $\tau\to\pi\etap\gamma\nu_\tau$ decays as background of the
non-radiative process. We find that the radiative processes are very important background for the non-radiative ones whenever the photon
escapes detection. Photons cannot be completely excluded since they need to be kept for the identification of the eta and eta' in the final
state. However, we find that making appropriate cuts in the energy of the photon, this process can be disregarded as an important background.

\end{abstract}

\section{Introduction}

In an effort to understand better the interactions between particles, especially of those that
interact strongly, T.D. Lee and C.N. Yang proposed a new operator \cite{Lee-Yang} to study the conservation
of its eigenvalues. This operator is defined as the product of a $\pi$ radians rotation around the $I_2$
direction in isospin space, times the charge conjugation operator
\begin{equation}
 G=Ce^{i\pi I_2}.
\end{equation}

They found that, for states with zero baryon number and strangeness, $N_B = S = 0$, the operator
$G$ has only two eigenvalues, $G = \pm1$. Therefore, for such states $G$ can be regarded as a parity
operator. And thus, hadronic currents with no change in strangeness, charm or bottomness,
$\Delta S = \Delta C = \Delta B = 0$, will have a definite $G$-parity. Depending on their $G$-parity, these
hadronic currents can be classified into first and second class as shown in Table \ref{GDef} .

\begin{table}[!ht]\label{GDef}\caption{Definition of first and second class (hadronic) currents according to their parity, spin and $G$-parity.}
    \centering
    \begin{tabular}{c c c c c}\hline\hline
    
             &$S$&$P$ &$V$&$A$\\ \hline
     $1^{st}$&$+1$&$-1$&$+1$&$-1$\\ 
     $2^{nd}$&$-1$&$+1$&$-1$&$+1$\\ \hline\hline
    
    \end{tabular}\\\hspace*{1ex}

   \end{table}
The Standard Model (SM) as constructed by Glashow, Weinberg and Salam \cite{SM} (and extended
to include quarks as done by Glashow, Iliopoulos and Maiani \cite{GIM}) involves only quark vector
currents with $G = +1$ for those with $\Delta S = \Delta C = \Delta B = 0$. Thus, a measurement of a Second
Class Current (SCC) could mean evidence for physics beyond the SM (BSM). However, such
signal could be also reproduced by an isospin-breaking current, which are present within the SM.
Therefore, SM induced SCC must be very well characterized in the search for BSM interactions.
Several channels for detection of SCC have been suggested, however the ones suggested by Leroy
and Pestieau \cite{SCC Leroy}, namely the $\tau^\pm\to\eta^{(\prime)}\pi^\pm\nu_\tau$ decays, are the cleanest channels in which to
search for genuine SCC. Since the experimental limits for $\eta$ \cite{Exp eta SCC} and $\eta'$ \cite{Exp etap SCC} for such process are
getting close to the theoretical prediction \cite{The SCC}, all possible background processes for the measurement of
these decays need to be in very good control in order for the experiments to give more precise and reliable results.
It so happens that processes in the SM with isospin breaking are suppressed by factors involving differences of light quark masses. In these decays, 
there is also a kinematical suppression factor (compared to the most common decays into pions), which in total gives
\begin{equation}
 \left(\frac{m_{\pi^0}}{m_\eta}\cdot\frac{m_d-m_u}{m_s}\right)^2\sim10^{-5}.
\end{equation}\\

On the other hand, the radiative process $\tau^-\to\eta^{(\prime)}\pi^-\gamma\nu_\tau$ involves the effective vertex $W^*\pi\eta^{(\prime)}\gamma$ 
which is not a $G$ eigenstate, since the photon does not have a definite isospin value. However, these decays will have an $\alpha_{EM}$ 
factor, which is of the order of the isospin suppression. 
Then, if in the radiative process the photon escapes detection, it could give a fake signal of a genuine SCC detection. Therefore, it becomes utterly 
necessary to give a prediction of the radiative decay to overcome the difficulty of its misidentification. To do so, we first estimate the 
Branching Fraction of the radiative process considering only the model dependent contribution stemming from the effective vertex 
$W^*\pi\eta^{(\prime)}\gamma$ using the Meson Dominance Model (MDM), which gives a simple description of the problem. We then give an  
estimation of the bremsstrahlung contribution to obtain a lower bound in the photon energy and safely neglect this contribution 
within Resonance Chiral Theory (\RCT) by neglecting the momentum of the photon in the form factors. Then we compute 
the form factors for the radiative process using {\RCT} to get the photon energy spectrum and compare it to that obtained through the MDM 
to give an upper bound in the photon energy so that the radiative decay can be safely neglected in the search for SCC. Other studies of 
background in the search for SCC through the $\tau\to\etap\pi\nu_\tau$ decays are being studied, see for example \cite{Tome}.

\section{Amplitude and form factors}

 In order to compute the required observables we need first the amplitude of the process. By choosing the four-momenta convention 
 $\tau(P)\to\pi^-(p)\eta^{(\prime)}(p_0)\nu_\tau(p')\gamma(k,\epsilon)$, one finds that the most general expression for this decay is
 \begin{equation}
  \mathcal{M}=\frac{eG_FV_{ud}^*}{\sqrt{2}}\epsilon^{*\mu}\left[\bar{u}(p')\gamma^\nu(1-\gamma_5)\left(V^{IB}_{\mu\nu}+V_{\mu\nu}+A_{\mu\nu}\right)u(P)\right],
 \end{equation}
where \begin{equation} 
       V^{IB}_{\mu\nu}=\frac{H_\nu(p,p_0)}{-2P\cdot k}(M_\tau+\slashed{P}-\slashed{k})\gamma_\mu+V^{SI}_{\mu\nu}
      \end{equation}
 is the bremsstrahlung contribution, {\it i.e.}, that is suppressed by the isospin and the $\alpha_{EM}$ factors. The remaining 
terms are structure dependent stemming from the $W^*\pi\eta\gamma$ effective vertex, which have no Dirac structure. In the previous equation, 
$H_\nu(p,p_0)=\langle\etap\pi^-|\bar{d}\gamma_\nu u|0\rangle$ is the hadronic current and $V^{SI}_{\mu\nu}$ is the corresponding part 
for the radiation off the $\pi^-$; their general form can be seen in \cite{Us2016}.\\

Now, gauge invariance can be applied to reduce the number of independent form factors, which gives the vector hadronic current $V_{\mu\nu}$
in terms of four form factors and, likewise, the axial hadronic current is given in terms of four axial form factors (using Schouten's identity),
\begin{subequations}
 \begin{align}
  V_{\mu\nu}=&v_1(p\cdot k\hspace{.5ex}g_{\mu\nu}-p_\mu k_\nu) + v_2(g_{\mu\nu}\hspace*{.5ex}p_0\cdot k-{p_0}_\mu k_\nu)\nonumber\\
  &+v_3(p_\mu\hspace*{.5ex} p_0\cdot k-{p_0}_\mu\hspace*{.5ex}p\cdot k)p_\nu\nonumber\\
  &+v_4(p_\mu\hspace*{.5ex} p_0\cdot k-{p_0}_\mu\hspace*{.5ex}p\cdot k){p_0}_\nu,\\
  A_{\mu\nu}=&i\varepsilon_{\mu\nu\rho\sigma}(a_1p_0^\rho k^\sigma+a_2k^\rho W^\sigma)\nonumber\\
  &+i\varepsilon_{\mu\rho\sigma\tau}k^\rho p^\sigma p_0^\tau\left[a_3W_\nu+a_4(p_0+k)_\nu\right],
 \end{align}
\end{subequations}
where $W=P-p'=p+p_0+k$. However, this decomposition is not unique and the form factors $v_i$ and $a_i$ will be determined from 
the specific model for the hadronic interactions. We will first compute them using the Meson Dominance Model to give an estimation of 
the SD contribution to the Branching Fraction.

\section{Estimation of the bremsstrahlung contribution}

 Although the bremsstrahlung contribution to the process will be suppressed by $\alpha_{EM}$ and the isospin factor, the infrared behavior of the 
 photon could at some point surpass such suppressions. Therefore it is necessary to estimate the effect of the $V^{IB}_{\mu\nu}$ term at low photon energies. 
 To do so we rely on Low's theorem \cite{Low}. This theorem proofs that the amplitude can be written as a power series in the photon's four momentum,
 \begin{equation}
  \mathcal{M}_\gamma=\frac{A}{k}+B+\mathcal{O}(k).
 \end{equation}
Here, $A$ and $B$ are two quantities that are given in terms of the non-radiative amplitude $\mathcal{M}_0$, in which the four momentum of the photon 
is neglected on the form factors. Since we are only interested in the low energy behavior of the photon's momentum the amplitude can be approximated by
\begin{equation}
 \mathcal{M}_\gamma=-e\mathcal{M}_0\left(\frac{P\cdot \epsilon}{P\cdot k}-\frac{p\cdot \epsilon}{p\cdot k}\right)+\cdots,
\end{equation}
being that higher orders in $k$ will emerge solely by keeping the energy dependence in the form factors of the $W^*\pi\etap$ effective vertex. The non-radiative 
amplitude can be computed from the expressions in ref. \cite{nonrad}.
\begin{figure}[!ht]
 \centering\includegraphics[scale=0.3,angle=-90]{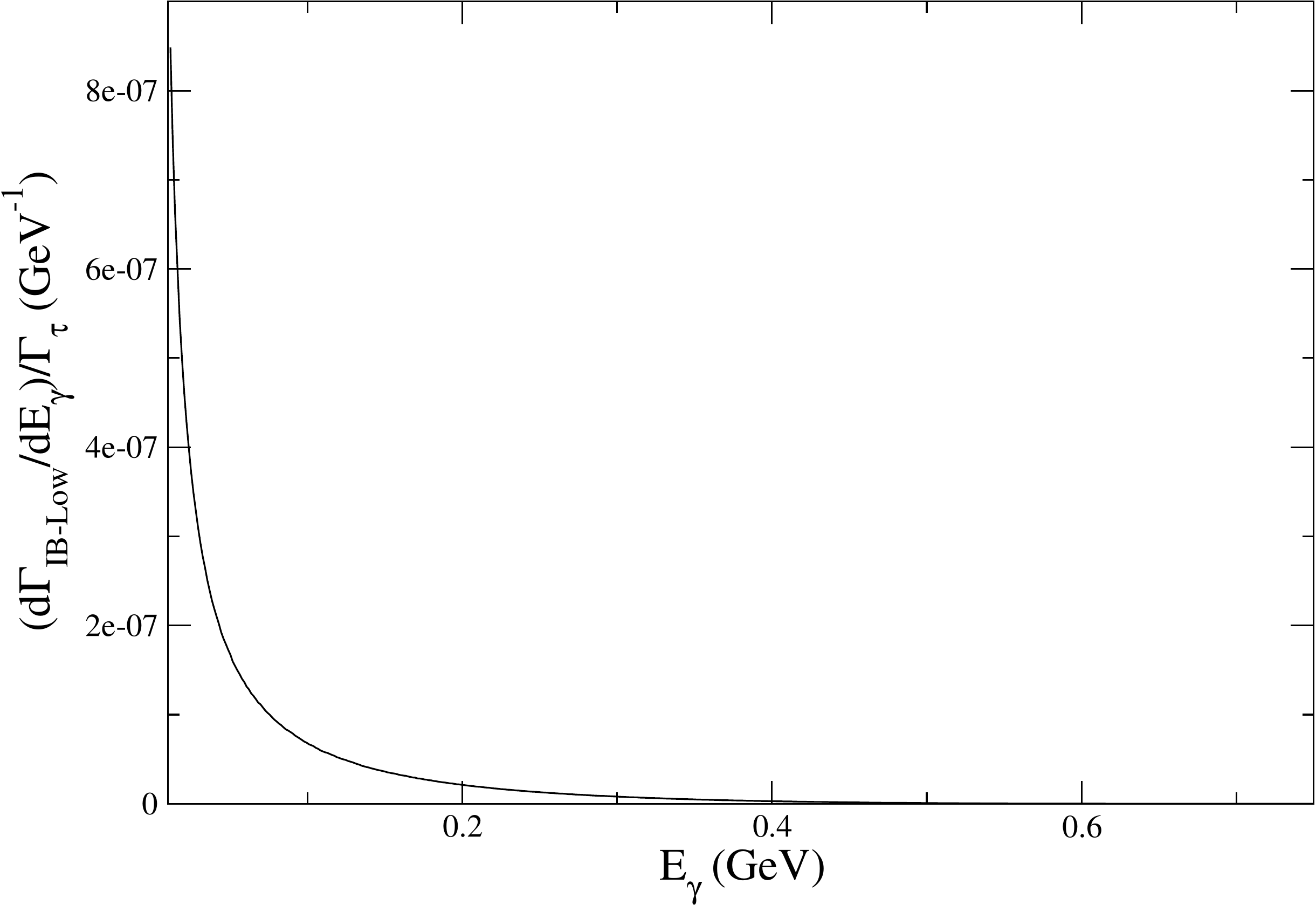}\hspace*{5ex}\includegraphics[scale=0.3, angle=-90]{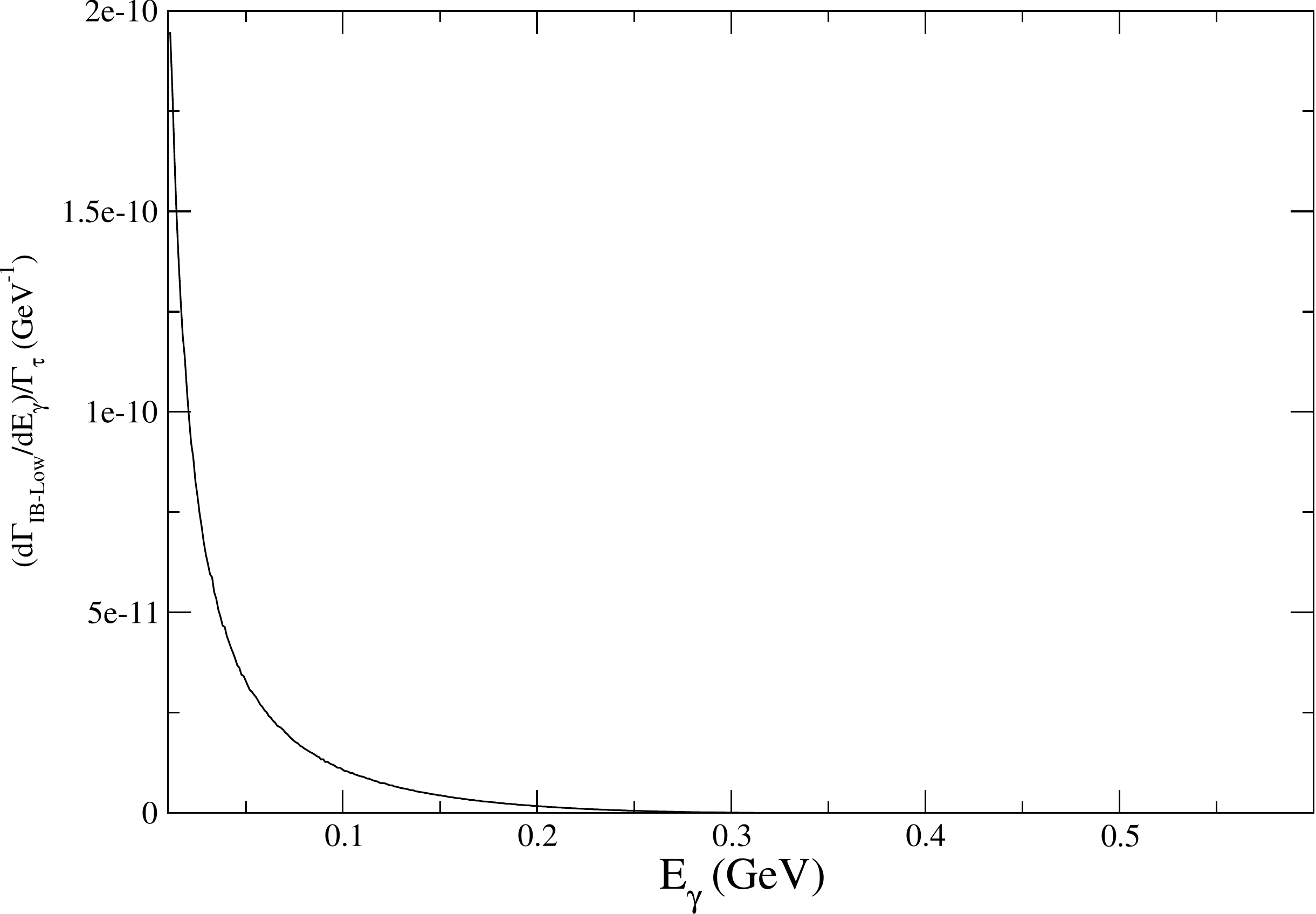}\caption{Photon energy spectra for the leading terms 
 of the bremsstrahlung amplitude for $\tau\to\pi\eta\gamma\nu_\tau$ (left) and $\tau\to\pi\eta'\gamma\nu_\tau$ (right).}\label{Bremss}
\end{figure}
Thus, by setting a cut for the energy of the photon of 10 MeV the Branching Ratios obtained are $\mathcal{B}(\tau\to\pi\eta\gamma\nu_\tau)\sim2.5\times10^{-8}$
and $\mathcal{B}(\tau\to\pi\eta'\gamma\nu_\tau)\sim4.6\times10^{-12}$, where the corresponding energy spectra are shown in Figures \ref{Bremss}. Therefore, applying 
this (realistic) bound on the photon's energy one will safely neglect the bremsstrahlung contribution.
\section{Form factors within the Meson Dominance Model}
\begin{figure}[!ht]
 \centering\includegraphics[scale=.6]{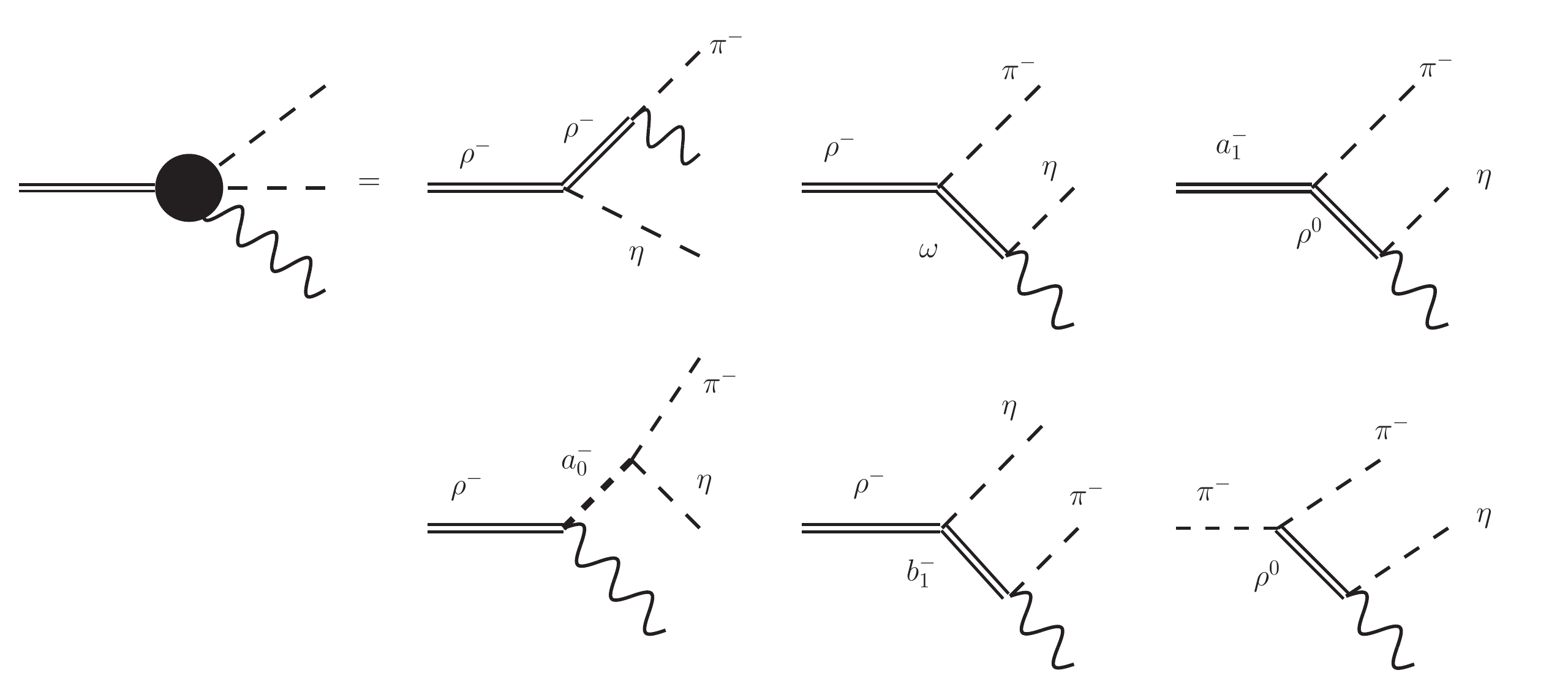}\caption{Contribution to the form factors in the Meson Dominance Model}\label{MDM}
\end{figure}

In the Meson Dominance Model (MDM), one assumes that the weak and electromagnetic couplings are dominated by the exchange of a few light mesons
and their excitations. This approach is useful provided one is able to determine the relevant couplings through data fitting from other independent sources 
or through model assumptions. The form factors are given by the diagrams of Figure \ref{MDM}, where the vertices are obtained with the rules given by
\begin{subequations}
 \begin{align}
  V'^\mu(r)\to V^\alpha(s)P(t):&ig_{V'VP}\varepsilon^{\mu\alpha\rho\sigma}s_\rho t_\sigma,\\
  V^\mu(r)\to \gamma^\alpha(s)P(t):&ig_{V\gamma P}\varepsilon^{\mu\alpha\rho\sigma}s_\rho t_\sigma,\\
  V^\mu(r)\to \gamma^\alpha(s)P(t):&ig_{V\gamma P}\varepsilon^{\mu\alpha\rho\sigma}s_\rho t_\sigma,\\
  A^\mu(r)\to V^\alpha(s)P(t):&ig_{VAP}(r\cdot s\hspace*{.5ex}g_{\mu\alpha}-r_\alpha s_\mu),\\
  V^\mu(r)\to \gamma^\alpha(s)S(t):&ig_{V\gamma S}(r\cdot s\hspace*{.5ex}g_{\mu\alpha}-r_\alpha s_\mu).
 \end{align}
\end{subequations}
The couplings are phenomenologically obtained from different independent decays. All such determinations of the coupling constants can be 
seen in ref. \cite{Us2016}. 
From all diagrams of Fig. \ref{MDM}, two are not taken into account. First, the contribution of the diagram with the $b_1$ exchange will be neglected due to 
the smallness of the Branching Ratios $\mathcal{B}(b_1\to\pi\gamma)=(1.6\pm0.4)\times 10^{-3}$ and $\mathcal{B}(b_1\to\rho\eta)<10\%$.
Second, the diagram with the pion pole is very suppressed since the pion is far off its mass shell. The second approximation cannot be taken 
within {\RCT}, since it breaks chiral symmetry. With this we find the expressions of the form factors to be
\begin{eqnarray}
 v_1^{MDM}&=&iC_\rho\left[-\frac{g_{\rho^-\rho^-\eta}g_{\rho^-\pi^-\gamma}}{D_\rho\left[(p+k)^2\right]}p\cdot p_0
 +\frac{g_{\rho^-\omega\pi^-}g_{\omega\eta\gamma}}{D_\omega\left[(p_0+k)^2\right]}p_0\cdot (p_0+k)
 +\frac{g_{\rho^-a_0^-\gamma}g_{a_0^-\pi^-\eta}}{D_{a_0}\left[(p+p_0)^2\right]}\right],\\
 v_2^{MDM}&=&iC_\rho\left[\frac{g_{\rho^-\rho^-\eta}g_{\rho^-\pi^-\gamma}}{D_\rho\left[(p+k)^2\right]}p\cdot (p+k)
 -\frac{g_{\rho^-\omega\pi^-}g_{\omega\eta\gamma}}{D_\omega\left[(p_0+k)^2\right]}p\cdot p_0
 +\frac{g_{\rho^-a_0^-\gamma}g_{a_0^-\pi^-\eta}}{D_{a_0}\left[(p+p_0)^2\right]}\right],\\
 v_3^{MDM}&=&iC_\rho\left[-\frac{g_{\rho^-\rho^-\eta}g_{\rho^-\pi^-\gamma}}{D_\rho\left[(p+k)^2\right]}\right],\\
 v_4^{MDM}&=&iC_\rho\left[\frac{g_{\rho^-\omega\pi^-}g_{\omega\eta\gamma}}{D_\omega\left[(p_0+k)^2\right]}\right],\\
 a_1^{MDM}&=&C_{a_1}\left[\frac{g_{\rho^-a_0^-\gamma}g_{a_0^-\pi^-\eta}}{D_{a_0}\left[(p+p_0)^2\right]}\right](p_0+k)\cdot W,\\
 a_2^{MDM}&=&0,\\
 a_3^{MDM}&=&0,\\
 a_4^{MDM}&=&-\frac{a_1^{MDM}}{(p_0+k)\cdot W}.
\end{eqnarray}
The definition of the constant $C_X$ is given in ref \cite{Us2016} and $D_X$ is the denominator of the propagator of resonance $X$ whose 
width may depend on the energy of the resonance.
\section{Form factors within the Resonance Chiral Theory}
Chiral Perturbation Theory ($\chi$PT) \cite{ChPT} is the effective field theory of low-energy Quantum Chromodynamics (QCD)
built upon its chiral symmetry. However, the validity of this theory stops when the lowest meson resonance, namely the $\rho$, becomes an active degree of freedom
($\sim700$ GeV). So, in order to extend the validity of the approach to higher energies the meson resonances must be included as active degrees of freedom. This 
is done by relying on the Large $N_C$ approximation of QCD in a manner compatible with chiral symmetry. This extension is called Resonance Chiral Theory ({\RCT} \cite{RCT})
and by including the lightest meson resonances it is able to extend its applicability to the GeV scale. It also incorporates information about high energy QCD by 
imposing relations from short distances to form factors, significantly reducing the number of free parameters of the model. It is worth to mention that no gauge symmetry 
principle is imposed to the resonances.\\
\begin{figure}[!ht]
 \centering\includegraphics[scale=0.5]{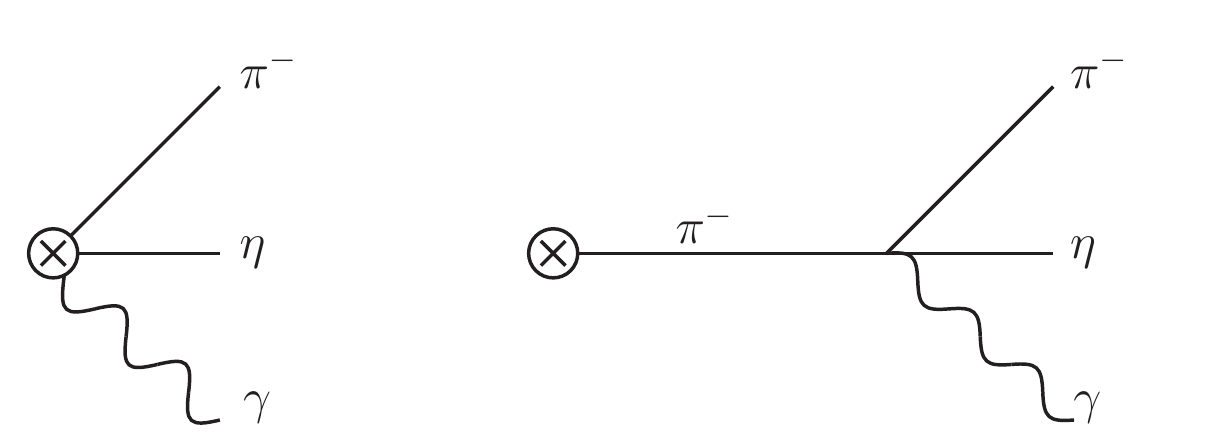}\caption{Contribution to the form factors from Chiral Perturbation Theory (no resonances).}\label{WZW}
\end{figure}
\begin{figure}[!ht]
 \centering\includegraphics[scale=0.44]{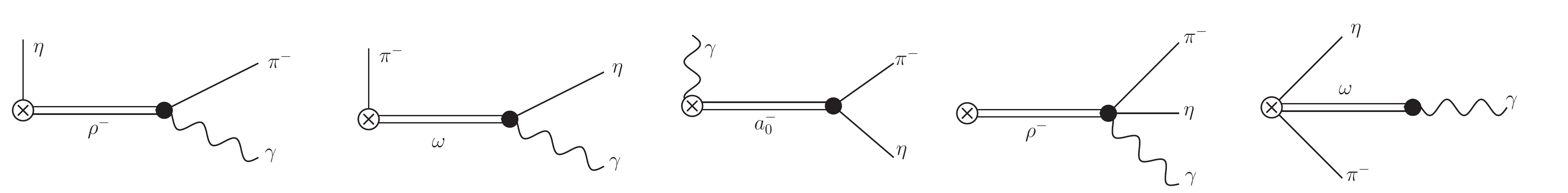}\\
 \centering\includegraphics[scale=0.44]{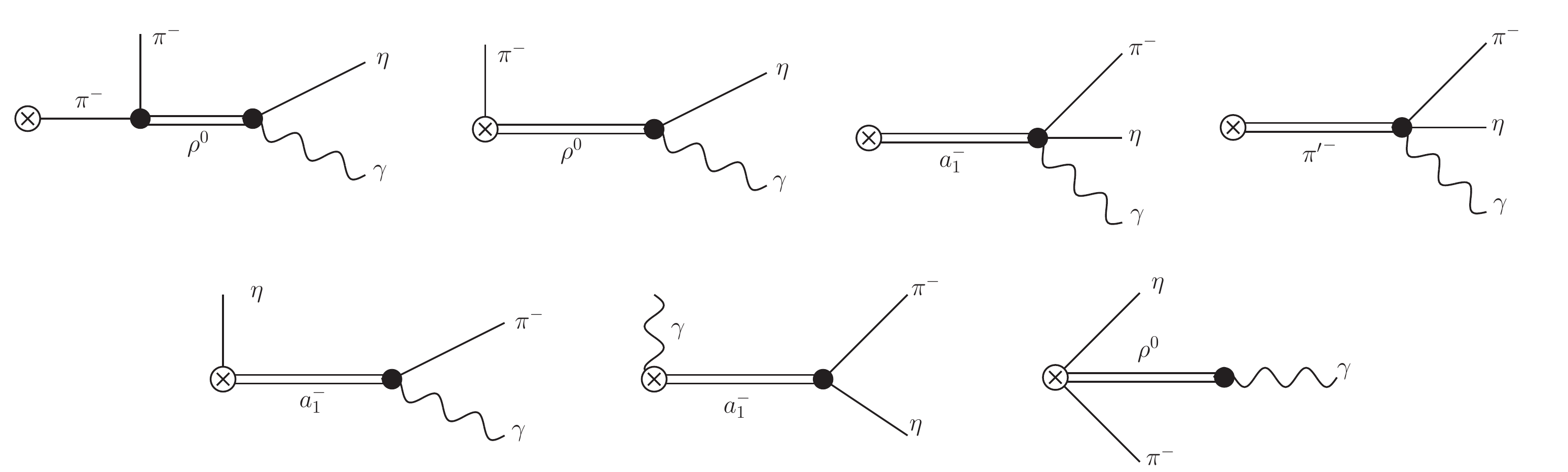}\caption{Contribution to the vector (first line) and axial (last two lines) form factors with one resonance exchange in \RCT.}\label{1R}
\end{figure}

Thus, we obtain that the contributions to the form factors are given by the diagrams shown in Figures \ref{WZW}, \ref{1R} and \ref{2R}. To compute every vertex of each 
diagram a larger base of operators must be considered than that in \cite{RCT}, since this gives only the even-intrinsic parity sector of the lowest chiral order. The 
operator base are obtained form references \cite{Kampf} and \cite{Cirigliano} for the odd-intrinsic and even-intrinsic parity sectors, respectively.
\begin{figure}[!ht]\label{2R}
 \centering\includegraphics[scale=0.33]{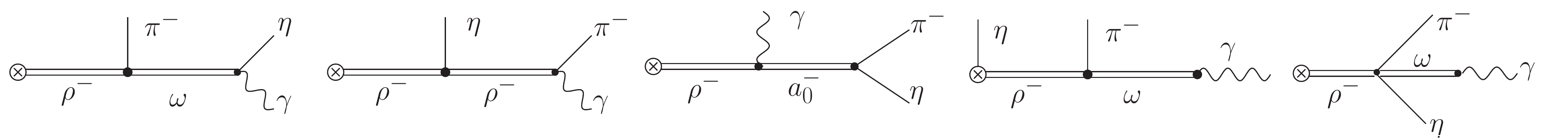}\\
 \centering\includegraphics[scale=0.33]{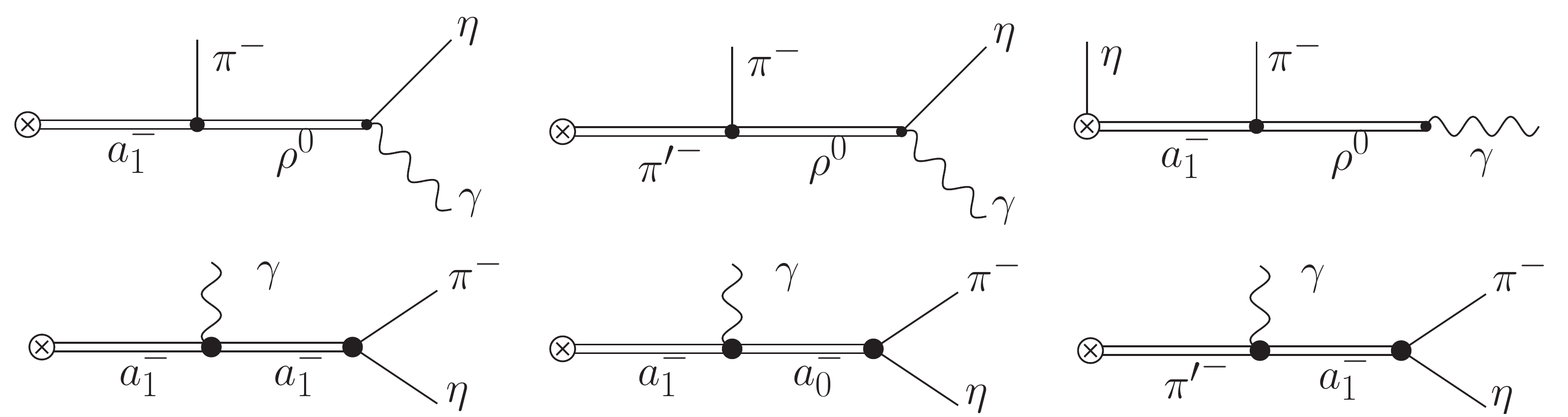}\caption{Contribution to the vector (first line) and axial (last two lines) form factors with two resonances exchange in \RCT.}\label{2R}
\end{figure}
Due to the extension of the form factors obtained with this model, they will not be shown here. The complete expressions are shown in ref \cite{Us2016}, along with the 
short-distance conditions \cite{Kampf,Juanjo} that fix some relations among the coupling constants and phenomenological determinations of given couplings. However, it must be noticed that not all the parameters of the theory can be determined 
through short distance or phenomenological relations, their absolute values are all bounded by some constraints stemming from the same order expansion constants in plain $\chi$PT
at NNLO \cite{Cirigliano}. Take, for example, the coupling strength of the terms in the lagrangian with only a vector resonance, $\lambda_i^V$ \cite{Kampf}. This 
coupling must be restricted by the relation
\begin{equation}
 \lambda^V_i\sim3C_i^R\frac{M_V^2}{F}\sim0.05 \text{ GeV}^{-1},
\end{equation}
and it can be taken as an upper bound for the absolute value of the coupling since the relation ($C_i^R$ is a NNLO constant of $\chi$PT) $C_i^R\sim\frac{1}{F^2(4\pi)^4}$ is linked to 
$L_i^R\sim\frac{1}{(4\pi)^2}\sim5\times10^{-3}$, which is the size of the largest NLO couplings, $|L_i^R|$. In a similar way, all bounds for the coupling
constants are obtained \cite{Us2016}. The uncertainties on the observables are obtained by randomly varying the values of the coupling constants\footnote{
Two different treatments have been followed in generating the values of the coupling constants: In the first, we generate randomly points within the bounds with uniform 
probability; in the second, we generate the points following a gaussian distribution, being the width equal to the bounds given. Here we will give the numerical values 
using the second method and present spectra using the first method; for the MDM parameters and the first method, see ref \cite{Thesis} for further discussion on 
the comparison between the two methods.} within the bounds given by the chiral couplings at NNLO.
\\

\section{Suppressing background in the search for SCC}
\begin{figure}[!ht]
 \centering\includegraphics[scale=0.3,angle=-90]{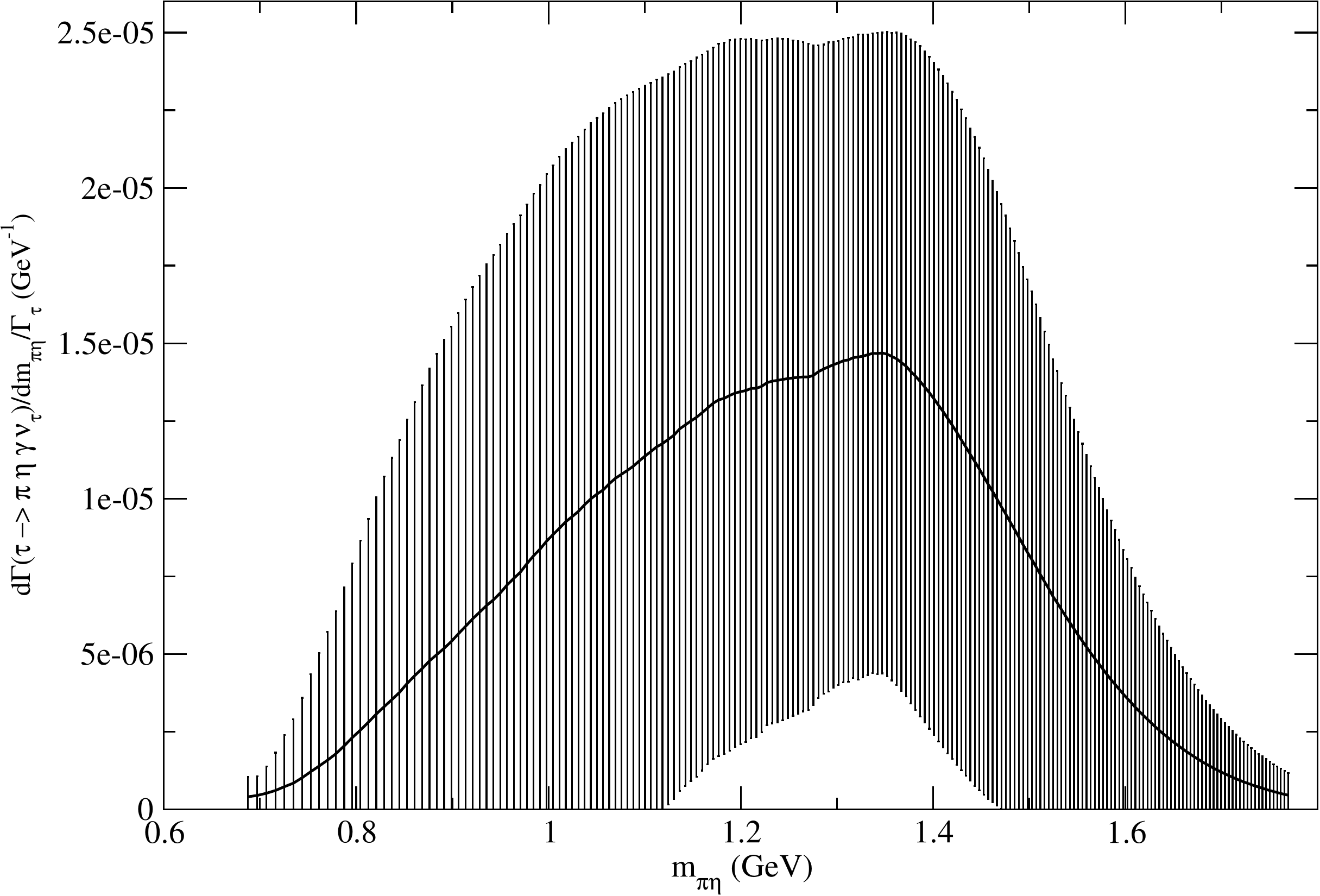}\hspace*{7ex}\includegraphics[scale=0.3, angle=-90]{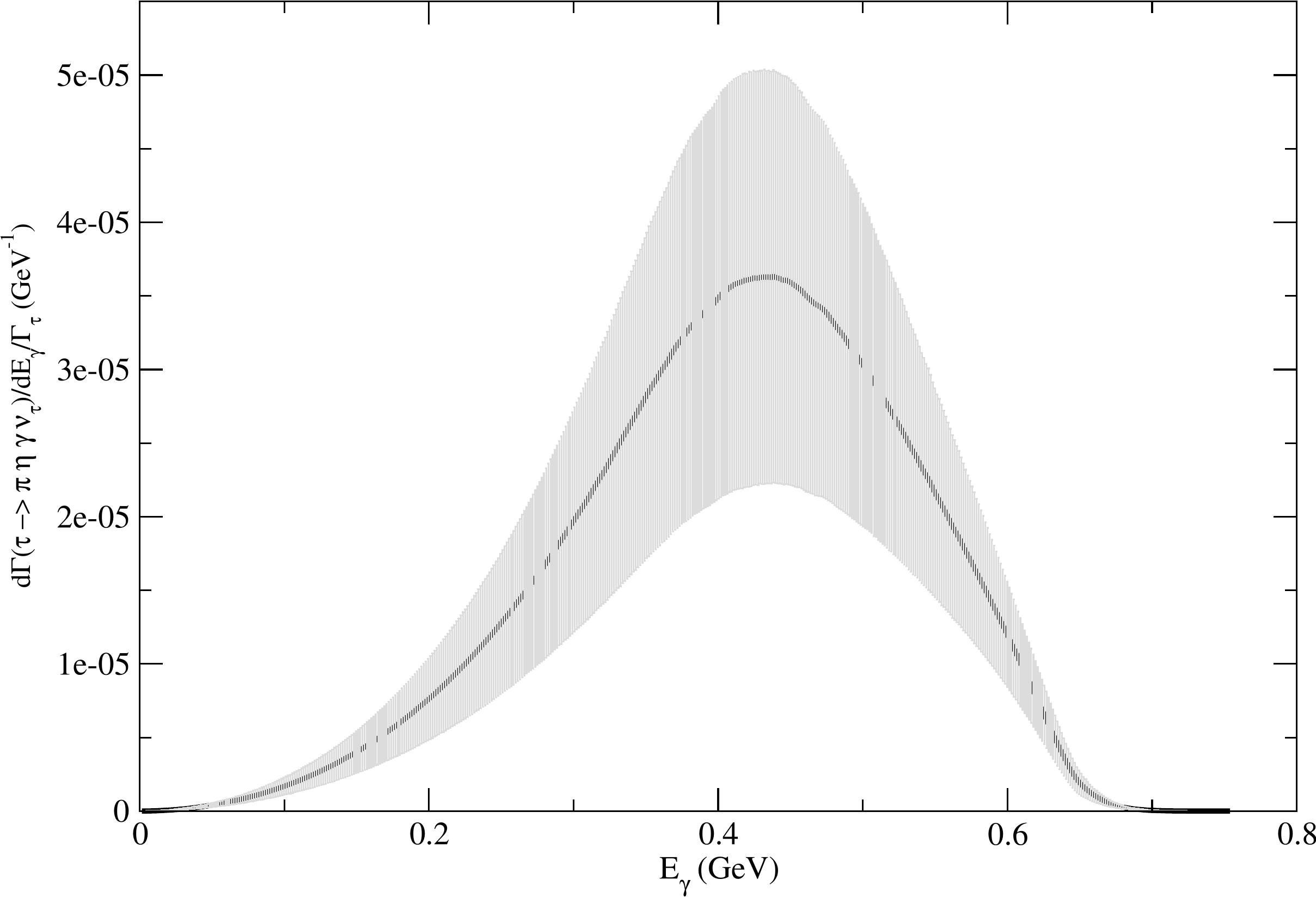}\caption{Normalized spectra for 
 $\mathcal{B}(\tau\to\pi\eta\gamma\nu_\tau)$ in MDM for the invariant mass of the $\pi^-\eta$ system (left) and photon energy (right).}\label{EgamMDM}
\end{figure}

\begin{figure}[!ht]
 \centering\includegraphics[scale=0.3,angle=-90]{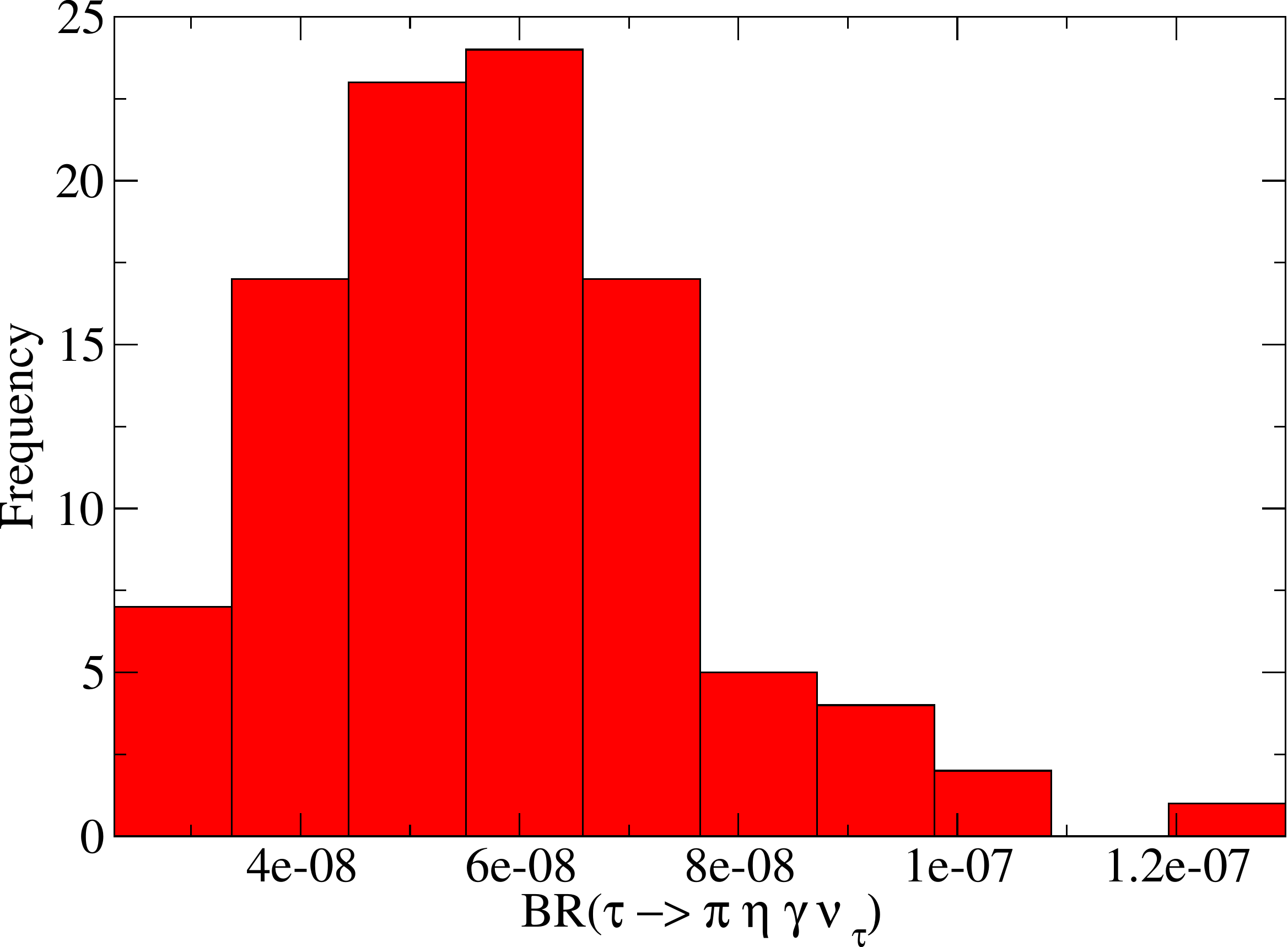}\caption{Histogram for $\mathcal{B}(\tau\to\pi\eta\gamma\nu_\tau)$ using 100 points in MDM parameter space
 rejecting photons with energies $E_\gamma>$100 MeV.}\label{BRcutMDM}
\end{figure}

In studying the $\tau\to\pi\etap\gamma\nu_\tau$ decays the question arises of whether one should consider the photon at all. Now, most of the ways in which the 
$\eta$ and $\eta'$ are detected involve photons, and therefore the analysis of the radiative process is utterly necessary to avoid a false signal of genuine SCC. 
The way to do so is by restricting the energy of the photon. So, to safely neglect the radiative process without affecting the detection of the non-radiative decay 
we must give a region of the phase space in which the former one is suppressed. Since the difference between both is the photon, it is natural to disregard the radiative 
process by discarding some values of the photon energy. The way to do this is, again, by relying on Low's theorem, which tells us that structure dependent contributions 
to the amplitude come from $\mathcal{O}(k)$ terms. Therefore, an upper bound to the energy of the photon must be applied to highly suppress the radiative process (along 
with the lower bound to neglect the bremsstrahlung contribution).\\
\begin{figure}[!ht]
 \centering\includegraphics[scale=0.3,angle=-90]{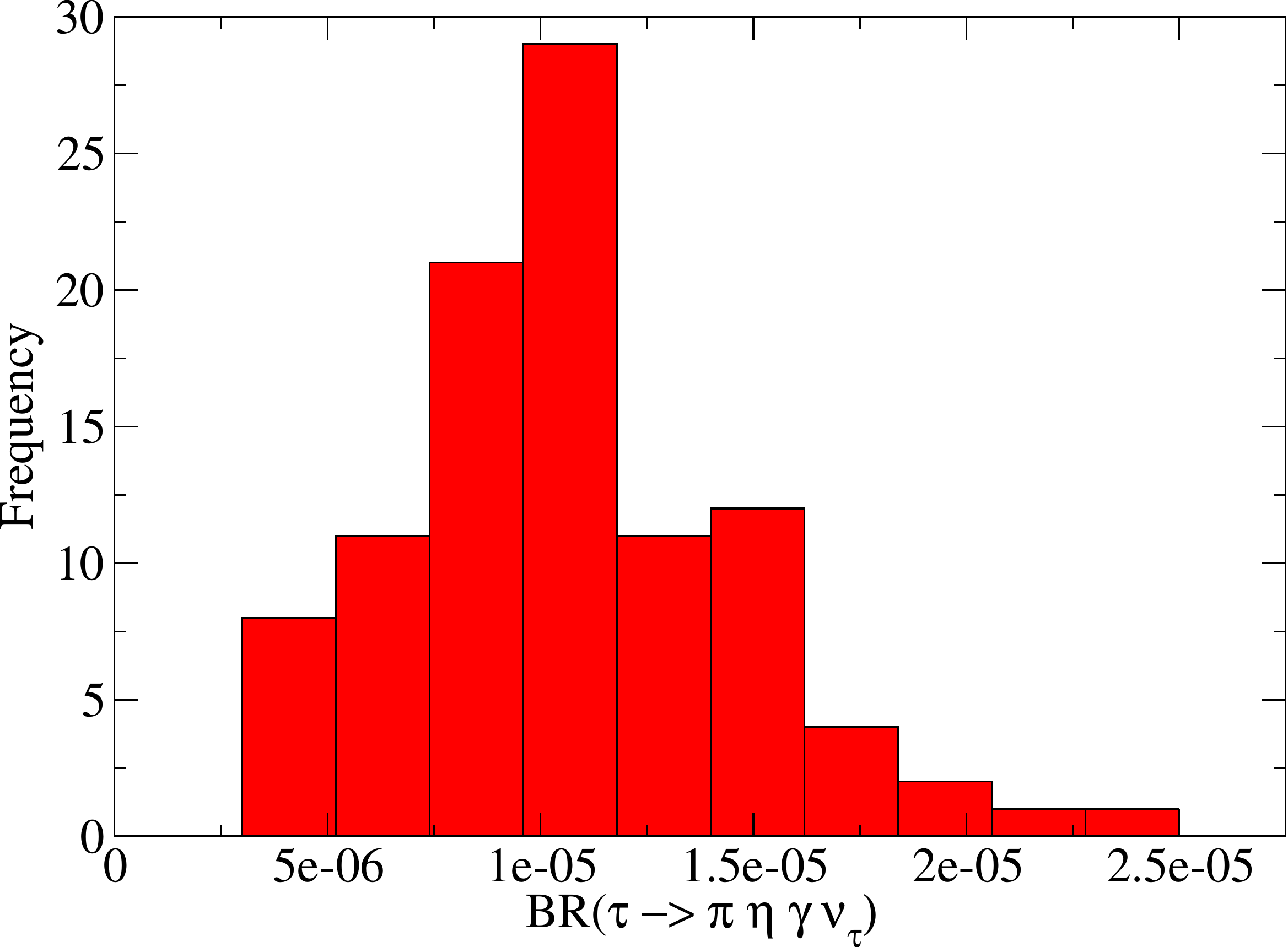}\hspace*{8ex}\includegraphics[scale=0.3, angle=-90]{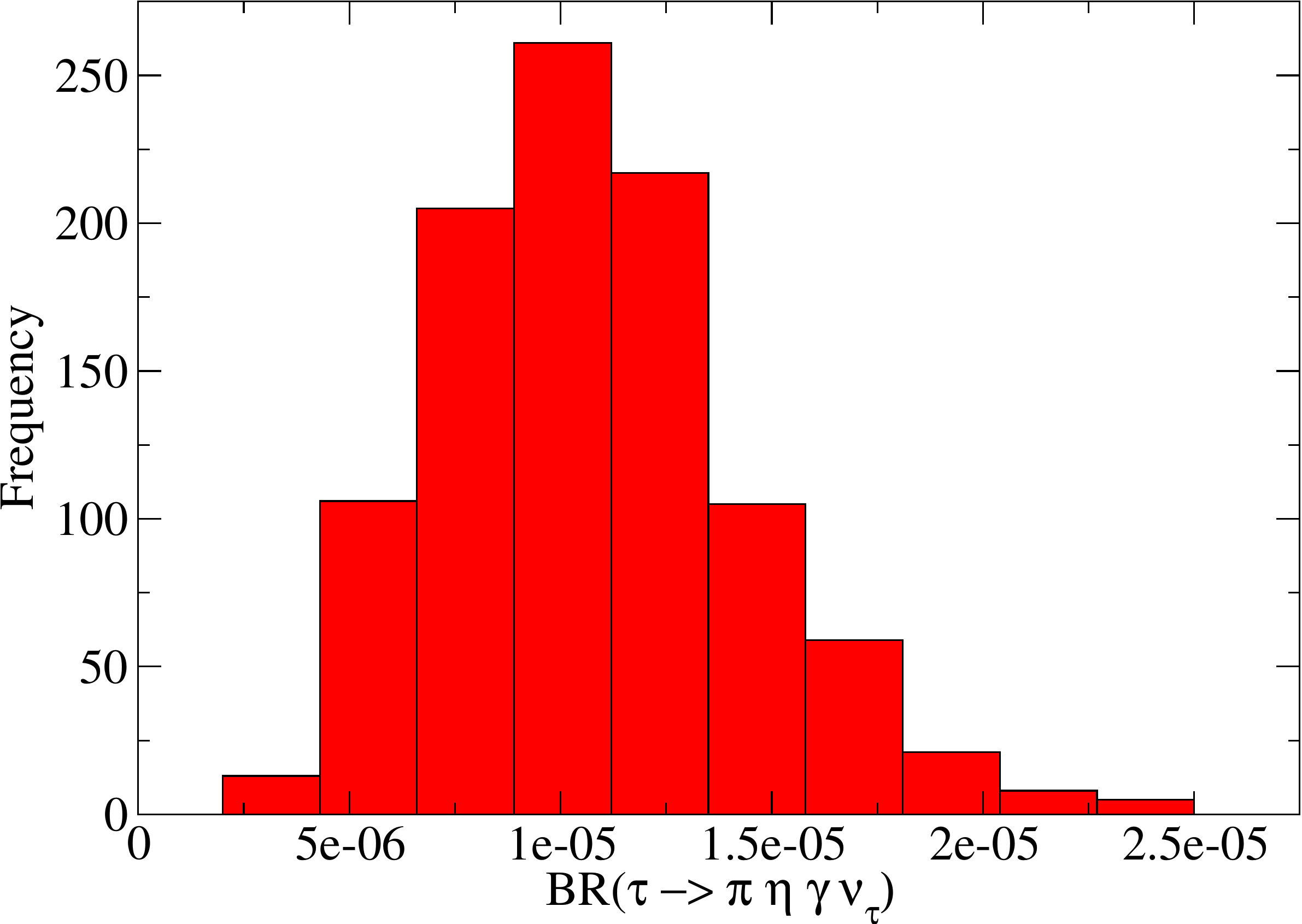}\caption{Histogram for 
 $\mathcal{B}(\tau\to\pi\eta\gamma\nu_\tau)$ for 100 (left) and 1000 (right) random points in parameter space in MDM.}\label{BRMDM}
\end{figure}
First, we will analyze the MDM prediction. What we find is that the Branching Fraction is, as supposed, of the order of the non-radiative decay $\left(\mathcal{B}
(\tau\to\pi\eta\nu_\tau)\sim10^{-5}\cite{nonrad}\right)$ as can be seen in Figs \ref{BRMDM}.
Also, from Figs \ref{EgamMDM} we can see that a good upper bound for the energy of the photon, $E_\gamma$ might be 100 MeV, which 
is verified by the results in Fig \ref{BRcutMDM}. 
It must be noticed that despite the peak around (1.25$\pm$0.05) GeV in the invariant mass $m_{\eta\pi}$, no marked dynamics is responsible for this effect. Therefore, by 
reevaluating the histogram with 1000 points in parameter space using the cut in $E_\gamma$ we obtain the bound $\mathcal{B}(\tau\to\pi\eta\gamma\nu_\tau)\le0.6\times10^{-7}$.\\

The procedure for the $\tau\to\pi\eta'\gamma\nu_\tau$ decay is completely analogous. The spectra and the histograms will not be shown here, but can be 
seen in ref \cite{Us2016}. By using a sample of 1000 points in parameter space and the same upper bound on $E_\gamma$ we find that 
$\mathcal{B}(\tau\to\pi\eta\gamma\nu_\tau)\le0.2\times10^{-8}$. It must be noticed that the MDM prediction is manly given by the $a_1^-$ exchange diagram; if all 
other contributions are neglected $\sim$80\% of the process is given by this contribution in the $\eta$ channel, while for the $\eta'$ it essentially saturates the 
amplitude.
\begin{figure}[!ht]
 \centering\includegraphics[scale=0.3,angle=-90]{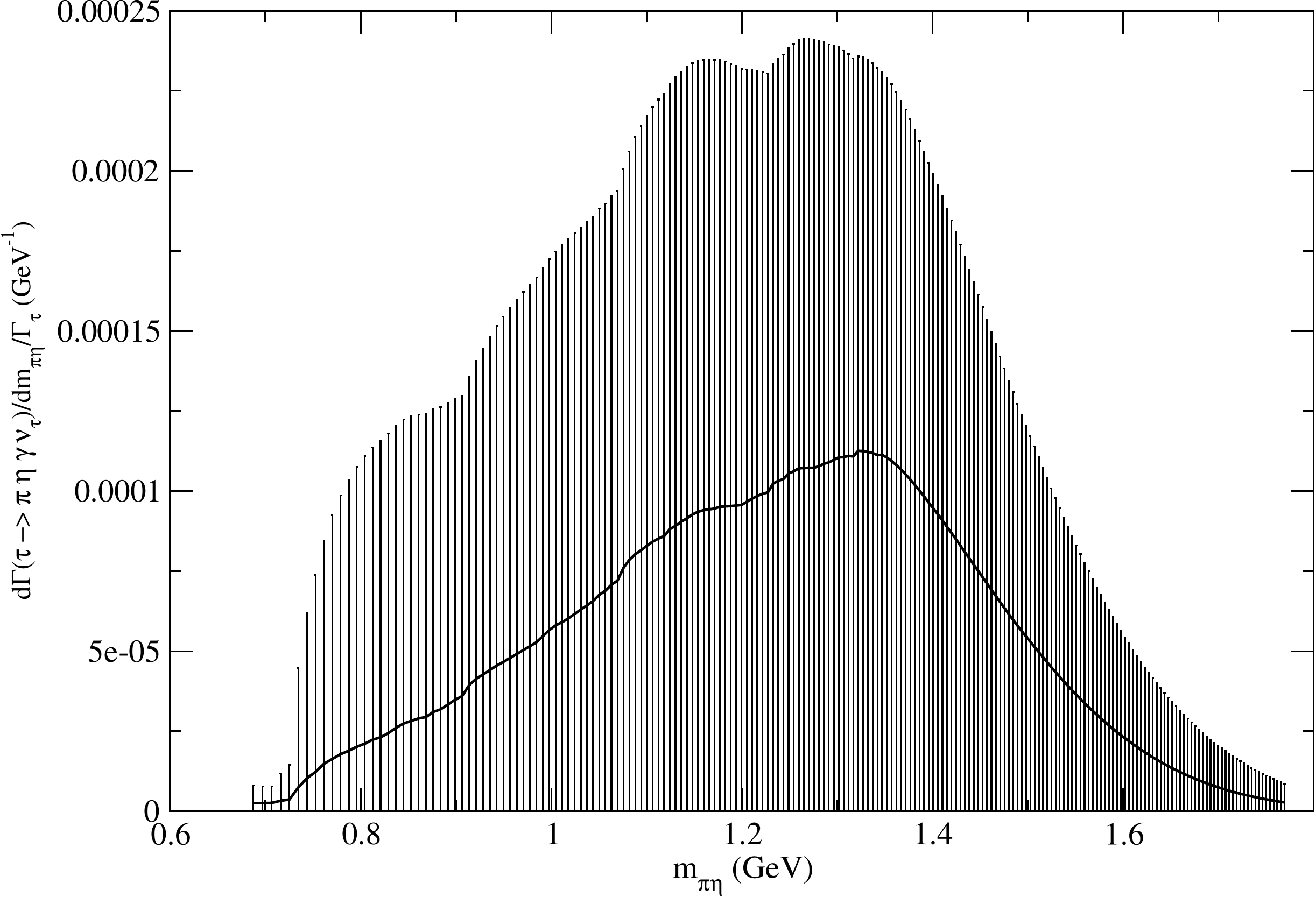}\hspace*{6ex}\includegraphics[scale=0.3, angle=-90]{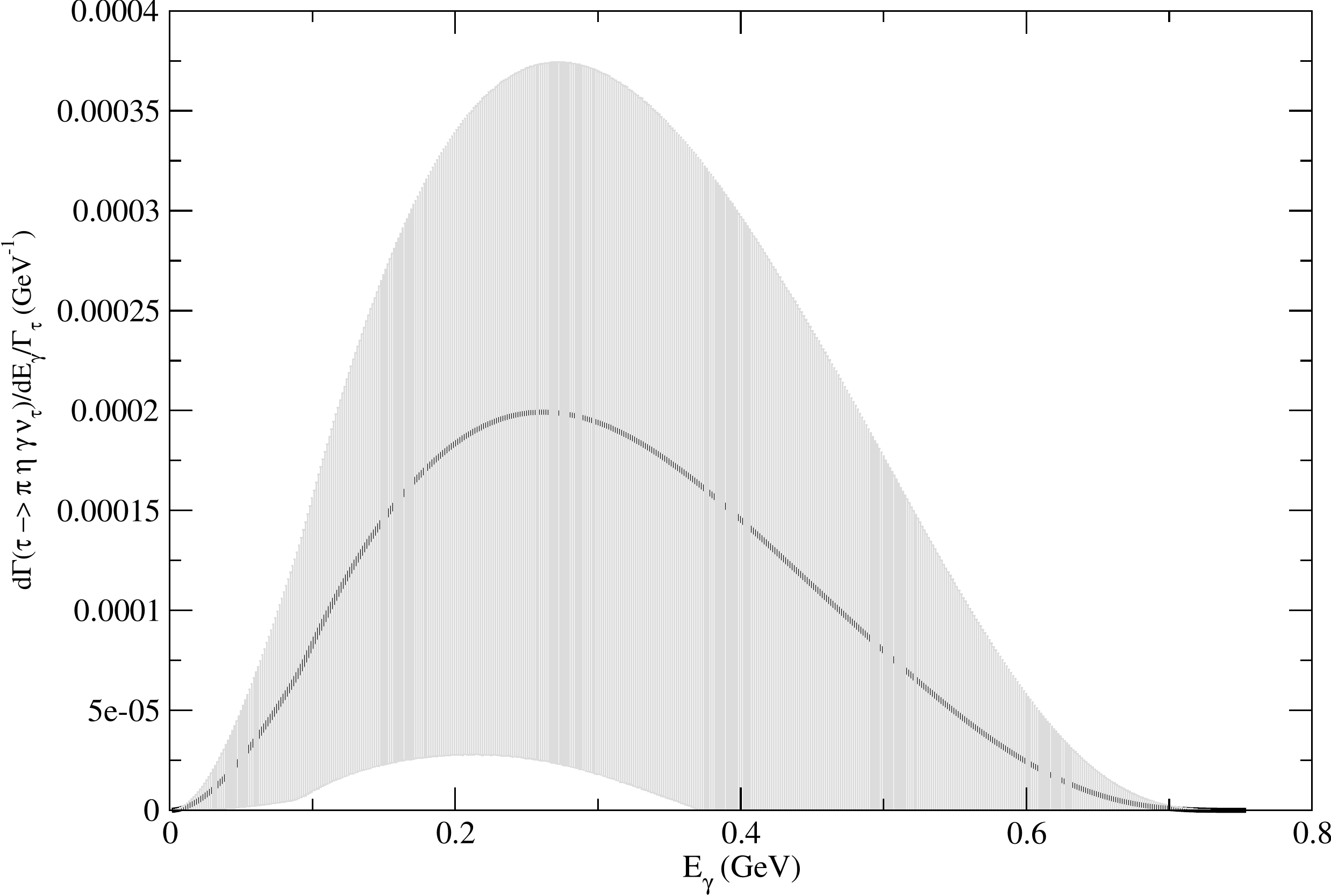}\caption{Normalized spectra for 
 $\mathcal{B}(\tau\to\pi\eta\gamma\nu_\tau)$ in {\RCT} for the invariant mass of the $\pi^-\eta$ system (left) and photon energy (right).}\label{EgamRChT}
\end{figure}\\
Using the form factors of {\RCT} one finds the spectra shown in Figs. \ref{EgamRChT}, which also points that a cut in $E_\gamma$ of 100 MeV will 
give a strong enough suppression and reproduce the behavior in the $[1.15,1.35]$ GeV region in the $m_{\eta\pi}$ spectrum obtained with MDM. 
A remarkable result comes about when the two-resonance-exchange diagrams are neglected, there is little difference between this and the full result,
as can be seen in Table \ref{Table1sigma}. This is an important feature, for our result can be implemented in TAUOLA without the two-resonance contributions, 
which will reduce computation time \cite{TAUOLA}.

\begin{table}[!ht]\caption{Branching fractions for different kinematical constraints and parameter space points.}\label{Table1sigma}
  \centering \begin{tabular}{ c c c c c}\hline\hline
    &$\begin{array}{c}\text{complete}\\\deceta\end{array}$&$\begin{array}{c}\text{without 2R}\\\deceta\end{array}$&$\begin{array}{c}\text{complete}\\\decetap\end{array}$&$\begin{array}{c}\text{without 2R}\\\decetap\end{array}$\\\hline
   100 points&$(2.3\pm0.9)\cdot10^{-5}$& $(2\pm2)10^{-5}$ &$(2.3\pm3.5)\cdot10^{-6}$&$(2.1\pm1.8)\cdot10^{-6}$ \\
   1000 points&$(3.0\pm0.6)\cdot10^{-5}$&$(2.3\pm0.5)\cdot10^{-5}$&$(2.2\pm0.4)\cdot10^{-6}$&$(2.0\pm0.4)\cdot10^{-6}$\\\hline
   $E_\gamma<100\MeV$&$(1.2\pm0.6)\cdot10^{-6}$&$(1.0\pm0.3)\cdot10^{-6}$&$(2\pm1)\cdot10^{-7}$&$(2\pm1)\cdot10^{-7}$\\\hline\hline
   \end{tabular}
 \end{table}

\section{Conclusions}

Belle-II will start collecting data in the very near future and because the experimental limits are now of the order of the theoretical 
prediction, this brings an excellent opportunity to look for SCC in $\tau$ decays. Encouraged by the prediction of MDM, we computed the 
Branching Fractions of the processes of interest in {\RCT}. Our results are shown in Table \ref{results} for the full phase space and 
after imposing the cut on the energy of the photon. We also show there the results of the induced SCC through isospin breaking from reference \cite{nonrad}.\\

\begin{table}[!ht]\caption{Our prediction of the Branching Ratios $\tau^-\to\pi^-\etap\gamma\nu_\tau$ decays without cuts and imposing a 
rejection of photons with $E_\gamma>100$ MeV.}\label{results}
 \centering
    \begin{tabular}{ c c c c} \hline \hline \hline
    Bkg & BR (no cuts) & BR ($E_\gamma^{\mathrm{cut}}>100$ MeV) & BR SCC signal\\\hline 
    $\eta$ & $(3.0\pm0.6)\cdot10^{-5}$ & $(1.2\pm0.6)\cdot10^{-6}$ & $\sim1.7\cdot10^{-5}$\\ 
    $\eta^\prime$ & $(2.2\pm0.4)\cdot10^{-6}$ & $(2\pm1)\cdot10^{-7}$ & $[10^{-7},10^{-6}]$\\ \hline \hline \hline
   \end{tabular}
\end{table}

The results of Table \ref{results} show that the $\deceta$ decay will have a negligible contribution to the background by imposing the bounds 
told on the energy of the photon; however, in the case of the $\decetap$ decay it is not the case, since the large uncertainty in the prediction 
of the non-radiative process shows that the radiative process might still be an important background in the search for genuine SCC in the $\eta'$ channel. 
The only way to surpass this problem would be to obtain a far more precise result of the isospin breaking prediction for the decay for the $\eta'$.\\

With this, we have pointed out for the first time the importance of the process studied as an important background on the search for SCC, 
where we found that the $G$-parity violation gives a comparable suppression factor to $\alpha_{EM}$, the electromagnetic one. We 
also found that the bremsstrahlung contribution is highly suppressed imposing a fairly lower bound on the energy of the photon $E_\gamma$.\\

We also found that the diagrams with two resonance exchange can be neglected in the computation of the form factors. This is a very useful 
result, since the form factors obtained through {\RCT} will be implemented in Monte Carlo generators. This will greatly reduce the time of 
computation and also will facilitate the implementation of the code by using the simplified expressions involving the no-resonance 
and the one-resonance-exchange diagrams. This can easily be seen in the complete expressions for the form factor available on ref \cite{Us2016}.

 \section*{Acknowledgments}
We have benefited from discussions on this topic with Jean Pestieau, Jorge Portol\'es, and Juan Jos\'e Sanz-Cillero. Financial support from projects 
296 (“Fronteras de la Ciencia”), 236394, 250628 (“Ciencia B\'asica”) and Sistema Nacional de Investigadores (SNI) (Conacyt, M\'exico) is acknowledged.

\end{document}